\documentclass[aps,prb,twocolumn,superscriptaddress,floatfix]{revtex4}

\usepackage{graphicx}
\usepackage{dcolumn}
\usepackage{bm}
\usepackage{times}
\usepackage{color}

\begin{document}
\bibliographystyle{apsrev}
\title{Unusual magnetotransport in holmium monoantimonide}
\author{Yi-Yan Wang}\thanks{These authors contributed equally to this paper}
\author{Lin-Lin Sun}\thanks{These authors contributed equally to this paper}
\author{Sheng Xu}
\author{Yuan Su}
\author{Tian-Long Xia}\email{tlxia@ruc.edu.cn}
\affiliation{Department of Physics, Renmin University of China, Beijing 100872, P. R. China}
\affiliation{Beijing Key Laboratory of Opto-electronic Functional Materials $\&$ Micro-nano Devices, Renmin University of China, Beijing 100872, P. R. China}
\date{\today}
\begin{abstract}
We report the magnetotransport properties of HoSb, a semimetal with antiferromagnetic ground state. HoSb shows extremely large magnetoresistance (XMR) and Shubnikov-de Haas (SdH) oscillation at low temperature and high magnetic field. Different from previous reports in other rare earth monopnictides, kinks in $\rho(B)$ and $\rho_{xy}(B)$ curves and the field dependent resistivity plateau are observed in HoSb, which result from the magnetic phase transitions. The fast Fourier transform analysis of the SdH oscillation reveals the split of Fermi surfaces induced by the nonsymmetric spin-orbit interaction. The Berry phase extracted from SdH oscillation indicates the possible nontrivial electronic structure of HoSb in the presence of magnetic field. The Hall measurements suggest that the XMR originates from the electron-hole compensation and high mobility.
\end{abstract}
\maketitle

\setlength{\parindent}{1em}
\section{Introduction}

In recent years, rare earth monopnictides LnX (Ln=rare earth elements; X=N, P, As, Sb, Bi) have been widely studied for their novel physical properties\cite{zeng2015topological,tafti2015resistivity,PhysRevLett.117.127204,Ban2017observation,PhysRevB.96.081112,PhysRevB.96.125112,sun2016large,PhysRevB.93.241106,PhysRevB.94.081108,PhysRevB.95.115140,nayak2017multiple,Singha2017Fermi,PhysRevMaterials.2.024201,PhysRevB.97.155153,tafti2016temperature,PhysRevB.93.235142,PhysRevB.94.165163,ghimire2016magnetotransport,Yu2017Magnetoresistance,pavlosiuk2016giant,PhysRevLett.117.267201,PhysRevB.96.075159,alidoust2016new,guo2017possible,PhysRevB.97.081108,PhysRevB.96.041120,PhysRevLett.120.086402,PhysRevB.93.205152,neupane2016observation,PhysRevB.97.115133,PhysRevB.97.085137,PhysRevB.96.125122,PhysRevB.96.035134,duan2018tunable,li2017predicted}. Remarkably, extremely large magnetoresistance (XMR) and resistivity plateau are often observed in these materials. The electron-hole compensation is a prevalent explanation for the origin of XMR\cite{ali2014large,PhysRevB.93.235142,Yu2017Magnetoresistance,PhysRevB.97.085137}. Conventional nonmagnetic metals only have a small magnetoresistance (MR). However, in the case of electron-hole compensation, high mobility of carriers will result in the emergence of XMR based on the semiclassical two-band model. Other mechanisms, such as the removed suppression of backscattering induced by the breaking of topological protection\cite{liang2015ultrahigh} or the change of spin texture\cite{PhysRevLett.115.166601}, have also been proposed to explain the XMR. The occurrence of resistivity plateau is universal in XMR materials\cite{tafti2015resistivity,PhysRevB.94.041103}, and it has been suggested to originate from the nearly invariable carrier concentration and mobility at low temperature\cite{PhysRevB.93.235142,PhysRevB.96.125112,PhysRevB.97.085137}. Usually, the field-dependent MR of nonmagnetic compensated semimetals exhibits unsaturated quadratic behavior. However, in antiferromagnets CeSb and NdSb, the field-induced metamagnetic transition from the antiferromagnetic (AFM) state to ferromagnetic (FM) state gives rise to unusual transport phenomena. Due to the influence of spin scattering, the $\rho$(\emph{B}) curve of NdSb shows a kink at high field and specific temperature range\cite{PhysRevB.93.205152}. The situation in CeSb is more complex, where the authors claim there exist several magnetic phases in the magnetic phase diagram, and the kinks are different at different temperatures\cite{PhysRevB.97.081108}. Consequently, it is interesting to study the transport properties in magnetic LnX materials.

Since LaX (X=N, P, As, Sb, Bi) are predicted to be topological semimetals or topological insulators\cite{zeng2015topological}, the topological property of rare earth monopnictides has attracted much attention. For the nonmagnetic materials LaSb/LaBi/YSb, only LaBi is considered to hold nontrivial band structure\cite{PhysRevB.95.115140,nayak2017multiple,PhysRevB.97.155153}, while LaSb\cite{PhysRevLett.117.127204,PhysRevB.96.041120} and YSb\cite{PhysRevLett.117.267201,Yu2017Magnetoresistance} are trivial semimetals. The topological transition from a trivial to a nontrivial phase in CeX (X=P, As, Sb, Bi) further emphasizes the importance of strong spin-orbit-coupling (SOC) effect\cite{PhysRevLett.120.086402}. For the paramagnetic (PM) TmSb and PrSb, topologically trivial characteristic of the band has been revealed by the trivial Berry phase or the study of electronic structure\cite{PhysRevB.97.085137,PhysRevB.96.125122}. The situation is different in CeSb and NdSb with magnetic phase transition. Although the band inversion is absent in the PM state of CeSb\cite{PhysRevB.97.081108,PhysRevB.96.041120}, the presence of Weyl fermion is suggested to be possible in the FM state of CeSb as supported by the observation of negative longitudinal MR and the electronic structure calculations\cite{guo2017possible}. The existence of Dirac semimetal state is suggested in the AFM state of NdSb, where the negative longitudinal MR has been observed and attributed to the chiral anomaly\cite{PhysRevB.97.115133}. The ferromagnetic GdSb is also predicted to hold Weyl fermions\cite{li2017predicted}. The topological properties of magnetic LnX materials still deserve further exploration.

HoSb possesses an AFM ground state. The magnetic structure of HoSb changes from MnO-type AFM arrangement to HoP-type arrangement, then to FM arrangement under the application of field\cite{PhysRev.131.922,brun1980quadropolar}. The magnetic transition makes HoSb a good platform to investigate the influence of magnetic interaction on the magnetotransport and topological properties. In this work, we grew the single crystals of HoSb and studied the magnetotransport properties. The observation of the kinks in $\rho(B)$ and $\rho_{xy}(B)$ curves and the field-dependent resistivity plateau indicate that the magnetotransport properties of HoSb is different from those of other rare earth monopnictides. The unusual properties are ascribed to the change of magnetic structure. At 2.65 K $\&$ 14 T, the transverse MR of HoSb reaches 2.42$\times$10$^4$$\%$. The XMR is suggested to originate from the electron-hole compensation and high mobility. The split of Fermi surfaces and the nontrivial Berry phase of the electron bands have been observed/extracted from the Shubnikov-de Haas (SdH) oscillation, which can be attributed to the nonsymmetric spin-orbit interaction and the possible nontrivial electronic structure of HoSb in the magnetic field, respectively.

\section{Experimental methods and crystal structure}

Single crystals of HoSb were grown from the Sb flux. Starting materials of Ho and Sb with a molar ratio of 1:6 were put into an alumina crucible and then sealed in a quartz tube. The quartz tube was heated to 1150$^{\circ}$C and held for 10 h. After that, it was cooled to 750$^{\circ}$C at a rate of 1$^{\circ}$C/h, where the excess Sb flux was removed with a centrifuge. Finally, the cubic crystals were obtained. The atomic proportion determined by energy dispersive x-ray spectroscopy (EDS, Oxford X-Max 50) was consistent with 1:1 for Ho:Sb. Single crystal and powder x-ray diffraction (XRD) patterns were obtained with a Bruker D8 Advance x-ray diffractometer using Cu K$_{\alpha}$ radiation. TOPAS-4.2 was employed for the refinement. Transport measurements were carried out on a Quantum Design physical property measurement system (QD PPMS-14 T). The magnetic properties were measured on a Quantum Design magnetic property measurement system (MPMS3).

The single crystal XRD pattern of HoSb presented in Fig. 1(a) reveals that the (\emph{0 0 2l}) plane is the surface of the crystal. HoSb crystallizes in a simple rock-salt structure, as shown in the inset of Fig. 1(a). Fig. 1(b) shows the powder XRD pattern of HoSb. The reflections are well indexed in the space group \emph{Fm}-3\emph{m} and the refined lattice parameter a is 6.13(1){\AA}. A selected cubic single crystal of HoSb is presented in the inset of Fig. 1(b).

\begin{figure}[htbp]
\centering
\includegraphics[width=0.48\textwidth]{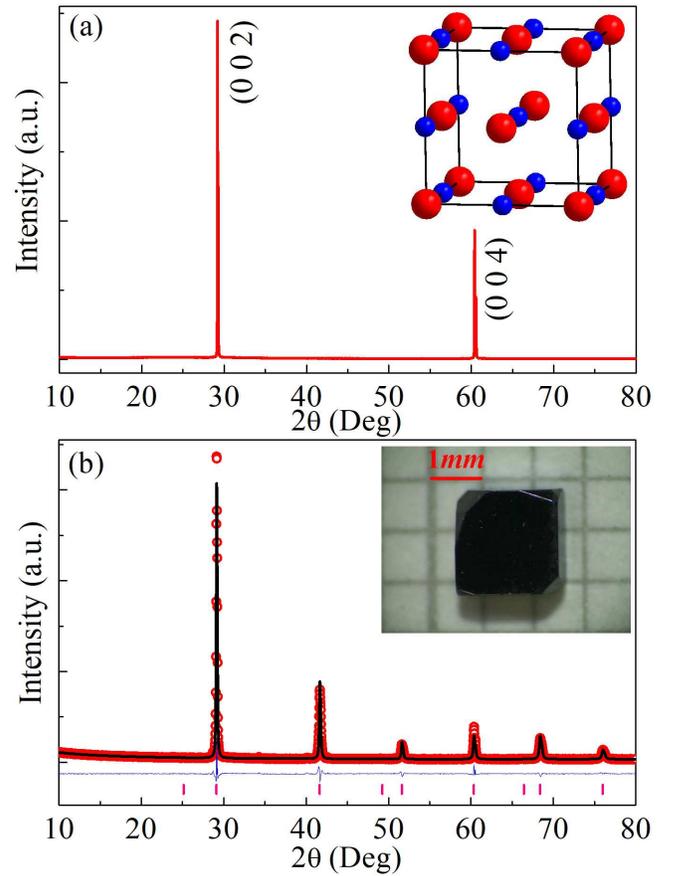}
\caption{(a) Single crystal XRD pattern of HoSb. The inset of (a) shows the NaCl-type crystal structure of HoSb. Red and blue balls represent Ho and Sb atoms, respectively. (b) Powder XRD pattern of HoSb refined using TOPAS. Red circle and black line represent the observed curve and calculated curve, respectively. Blue line denotes the difference curve. The vertical pink lines show the positions of Bragg peaks of HoSb. Inset presents a selected single crystal of HoSb.}
\end{figure}

\section{Results}

Fig. 2(a) shows the temperature-dependent resistivity of HoSb at various magnetic fields. The applied magnetic field is along [001] direction, perpendicular to the direction of the electric current. The inset shows the zero field resistivity in the low-temperature region. With decreasing temperature, the zero field resistivity decreases above the N$\acute{e}$el temperature $T_N$ = 5.7 K while below $T_N$ it drops suddenly. The anomaly observed around $T_N$ corresponds to the transition from PM to AFM, which is suppressed gradually under field. When a moderate field is applied, an upturn appears in the resistivity curve as the temperature decreases. Apart from the upturn, a resistivity plateau is observed with the further decrease of the temperature, which can be seen clearly from Fig. 2(b) where the temperature is plotted in a logarithmic scale. Similar behavior has also been observed in other XMR materials, such as the isostructural compounds La(Sb/Bi)\cite{tafti2015resistivity,sun2016large} and YSb\cite{Yu2017Magnetoresistance}. Remarkably, the temperature where the plateau starts increases with increasing field in HoSb. Fig. 2(c) plots the temperature dependence of \emph{$\partial\rho/\partial T$} at different fields. Two characteristic temperatures \emph{T$_m$} and \emph{T$_i$} can be identified from the \emph{$\partial\rho/\partial T$} curves, as shown in the inset of Fig. 2(c). \emph{T$_m$} is defined as the temperature where the sign of \emph{$\partial\rho/\partial T$} changes from positive to negative and \emph{T$_i$} is defined as the temperature where the valley appears. The resistivity reaches the minimum at \emph{T$_m$} and a resistivity plateau emerges below \emph{T$_i$}. Fig. 2(d) displays the \emph{T$_m$} and \emph{T$_i$} derived from Fig. 2(c) as a function of field. \emph{T$_m$} and \emph{T$_i$} both increase with increasing field. The increasing \emph{T$_i$} differs greatly from the nearly field-independent \emph{T$_i$} in other XMR materials\cite{PhysRevB.94.041103,tafti2015resistivity,sun2016large,Lv2016Extremely}.

\begin{figure}[htbp]
\centering
\includegraphics[width=0.48\textwidth]{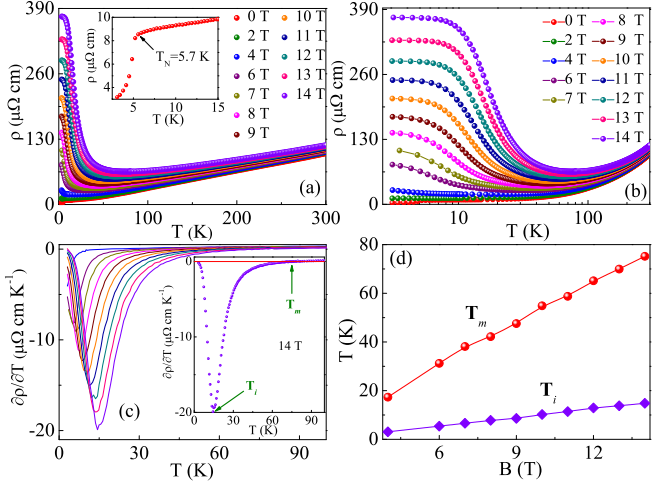}
\caption{(a) Temperature dependence of resistivity of HoSb (Sample 1, RRR = 32) at different applied magnetic fields. The inset displays zero-field resistivity in the low-temperature region. (b) Temperature-dependent resistivity with the transverse coordinate in the form of logarithm. (c) \emph{$\partial\rho/\partial T$} as a function of temperature at various fields from 4 T to 14 T. Inset shows the definitions of \emph{T$_i$} and \emph{T$_m$}. (d) \emph{T$_m$} and \emph{T$_i$} plotted as a function of field.}
\end{figure}

Magnetoresistance describes the change of the resistivity under magnetic field and can be obtained by the formula $MR = (\rho(B)-\rho(0))/\rho(0)$. Figure. 3(a) shows the transverse MR of HoSb plotted as a function of field at various temperatures. The value of MR reaches 2.42$\times$10$^4$$\%$ at 2.65 K and 14 T without any sign of saturation and decreases with increasing temperature. In addition, obvious SdH oscillation is also observed at low temperature and high field. The inset of Fig. 3(a) shows the oscillatory component of resistivity ($\Delta\rho_{xx}=\rho_{xx}-<\rho_{xx}>$) after subtracting a smooth background. Figure. 3(b) plots the fast Fourier transform (FFT) spectra of the oscillation, which reveal the frequencies. Considering the analysis of angle-dependent SdH oscillation (not presented here) and previous studies on the FSs of rare earth monopnictides, we identified the frequencies and drew the possible projection of FSs along $k_x$ (the inset of Fig. 3(b)). There are several pairs of peaks in the FFT spectra. Since the frequency $F$ is proportional to the extremal cross-sectional area $A$ of FS normal to the field according to the Onsager relation $F=(\phi_0/2\pi^2)A=(\hbar/2\pi e)A$, the feature that the frequencies appear in pairs indicates the split of FSs. The split of FSs has also been observed in TmSb and attributed to the nonsymmetric spin-orbit interaction\cite{PhysRevB.97.085137}. In the measurement, the current \emph{I} and field \emph{B} are parallel to \emph{x} axis and \emph{z} axis, respectively. The elliptical FSs can be divided into three categories: $\alpha^{\prime}$,
$\alpha^{\prime\prime}$ and $\alpha^{\prime\prime\prime}$. The absence of the frequency from $\alpha^{\prime\prime\prime}$ may be related to the low mobility along the long axis of the elliptical FSs, which has been derived in YSb and LaSb\cite{PhysRevB.96.075159,PhysRevB.96.125112}.

\begin{figure}[htbp]
\centering
\includegraphics[width=0.48\textwidth]{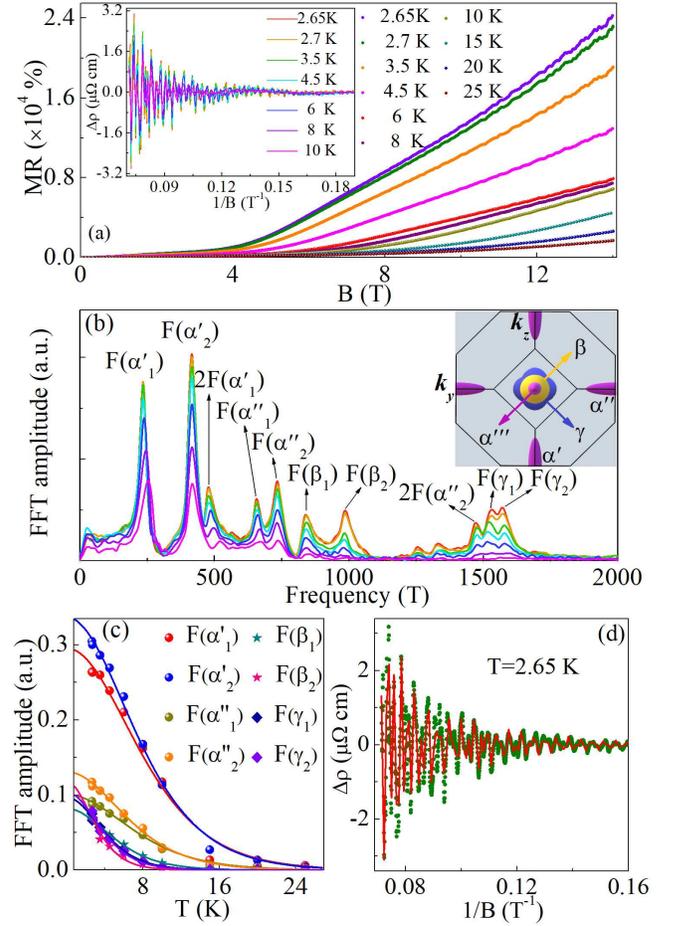}
\caption{(a) Magnetic field dependence of MR (Sample 2, RRR = 41) at different temperatures. Inset: The oscillatory part of resistivity plotted as a function of 1/\emph{B}. (b) The Fast Fourier transform (FFT) spectra of the SdH oscillation. Inset: The possible projection of Fermi surfaces of HoSb derived from the SdH oscillation. (c) Temperature-dependent FFT amplitudes of the frequencies, fitted with the thermal factor in Lifshitz-Kosevich (LK) formula. (d) The SdH oscillation at 2.65 K. The red solid line is the fitting result using the multiband LK formula.}
\end{figure}

In general, the Lifshitz-Kosevich (LK) formula\cite{shoenberg1984magnetic,PhysRevMaterials.2.021201}
\begin{equation}\label{equ1}
\Delta\rho\propto\frac{\lambda T}{sinh(\lambda T)}e^{-\lambda
T_D}cos[2\pi\times(\frac{F}{B}-\frac{1}{2}+\beta+\delta)]
\end{equation}
is employed to describe the amplitude of SdH oscillation. In the formula, $R_T=(\lambda T)/sinh(\lambda T)$ is the thermal factor, where $\lambda= (2\pi^2k_{B}m^*)/(\hbar eB)$ ($m^*$ and $k_B$ are the effective mass of carrier and the Boltzmann constant, respectively). $T_D$ is the Dingle temperature. $2\pi \beta$ is the Berry phase. $\delta$ is a phase shift. For the 2D system, the value of $\delta$ is 0. For the 3D system, $\delta$ takes $+1/8$ (hole pocket) or $-1/8$ (electron pocket)\cite{shoenberg1984magnetic}. Figure. 3(c) plots the temperature dependent FFT amplitude. The solid lines are fittings using the thermal factor $R_T$. As shown in Table I, the obtained effective masses are comparable with that of LaSb and TmSb\cite{tafti2015resistivity,PhysRevB.97.085137}. Berry phase can be extracted from the SdH oscillation and used to estimate the topological property of materials roughly. For such a multifrequency oscillation, it is difficult to map the Landau level index fan diagram, so we use the multiband LK formula to fit the oscillation pattern (Fig. 3(d)). The obtained Berry phase and Dingle temperature are listed in Table I. The values of Berry phase corresponding to the electron pocket (including the frequencies of $\alpha^{\prime}_1$, $\alpha^{\prime}_2$, $\alpha^{\prime\prime}_1$ and $\alpha^{\prime\prime}_2$) are close to the nontrivial value $\pi$, while the others are far away from $\pi$. However, a significant gap at \emph{X} point in HoSb has been suggested by the ARPES experiments performed in the absence of magnetic field\cite{PhysRevB.96.035134}. It is possible that the Weyl fermions may exist in the FM state of HoSb. The similar situation has also been suggested in CeSb\cite{guo2017possible} by the transport evidence and in GdSb\cite{li2017predicted} by the first-principles calculations. In addition, it should be noted that the $\pi$ Berry phase cannot be viewed as a smoking gun of topologically nontrivial material\cite{PhysRevX.8.011027}. Further study is needed to check the topological property of HoSb in the presence of the magnetic field.

\begin{table}
  \centering
  \caption{Parameters obtained from the analysis of SdH oscillation. $F$, oscillation frequency; A, extremal cross-sectional area of FS normal to the field; $k_F$, Fermi vector; $m^*$, effective mass of carrier; $T_D$, Dingle temperature; $2\pi\beta$, Berry phase ($\delta$ has been included).}
  \label{oscillations}
  \begin{tabular}{ccccccc}
    \hline\hline
    & $F$ (T) & A (${\AA}^{-2}$) & $k_F$ (${\AA}^{-1}$) & $m^*/m_e$ & $T_D$ (K) & $2\pi\beta$ \\
    \hline
$\alpha^{\prime}_{1}$ & 235.3 & 0.022 & 0.085 & 0.156 & 13.2 & 1.05$\pi$ \\
$\alpha^{\prime}_{2}$ & 416.3 & 0.040 & 0.112 & 0.164 & 17.5 & 0.89$\pi$ \\
$\alpha^{\prime\prime}_{1}$ & 657.2 & 0.063 & 0.141 & 0.173 & 29.3 & 1.17$\pi$ \\
$\alpha^{\prime\prime}_{2}$ & 733.7 & 0.070 & 0.149 & 0.191 & 27.4 & 0.83$\pi$ \\
$\beta_{1}$ & 839.9 & 0.080 & 0.160 & 0.250 & 17.9 & 0.14$\pi$ \\
$\beta_{2}$ & 986.5 & 0.094 & 0.173 & 0.414 & 14.7 & 0.48$\pi$ \\
$\gamma_{1}$ & 1532.0 & 0.146 & 0.216 & 0.311 & 20.6 & 0.09$\pi$ \\
$\gamma_{2}$ & 1570.7 & 0.150 & 0.218 & 0.336 & 18.3 & 0.34$\pi$ \\
    \hline\hline
  \end{tabular}
\end{table}

\begin{figure*}[htbp]
\centering
\includegraphics[width=\textwidth]{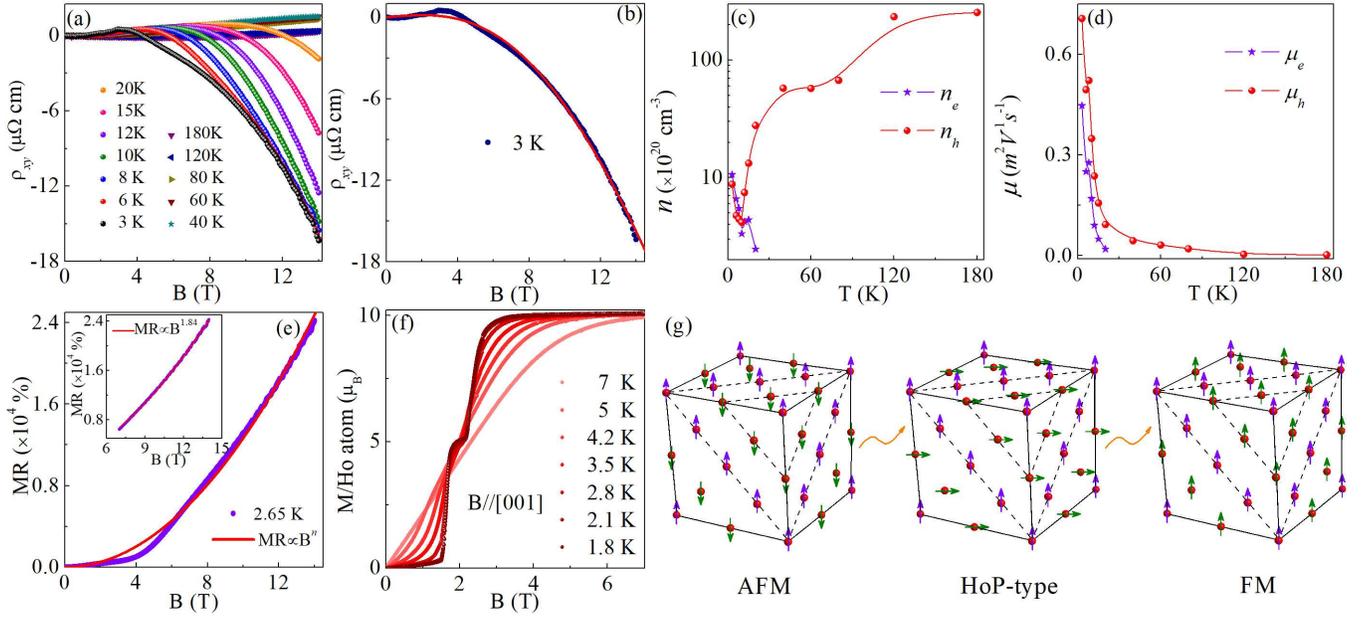}
\caption{(a) The magnetic field dependence of Hall resistivity (Sample 3) at various temperatures. (b) The Hall resistivity at 3 K and the fit (red solid line) using two band model. (c), (d) The temperature dependent concentration and mobility. The data at high temperatures was analyzed by single band model. (e) The fit to the field dependent MR (Sample 2) at 2.65 K using MR$\propto$\emph{B}$^n$. Inset: The fit only to the data in high field (7 T-14 T). (f) Magnetization plotted as a function of field with the field parallel to [001]. (g) The change of magnetic structure with the increase of field.}
\end{figure*}

Hall measurements are performed to reveal the origin of XMR in HoSb. As shown in Fig. 4(a), the field dependent Hall resistivity $\rho_{xy}=[\rho_{xy}(+B)-\rho_{xy}(-B)]/2$ exhibits nonlinear behavior, indicating the coexistence of electron and hole. With the increase of temperature, the dominant carrier changes from electron to hole. Unlike the Hall resistivity of usual two-component systems, there is a kink in the $\rho_{xy}(B)$ of HoSb, which will be discussed later. Figure. 4(b) shows the $\rho_{xy}(B)$ at 3 K. Apart from the imperfect part induced by the kink, the curve can be well fitted by the semiclassical two-band model,
\begin{equation}\label{equ2}
\rho_{xy}=\frac{B}{e}\frac{(n_h \mu_h^2-n_e \mu_e^2)+(n_h-n_e)(\mu_h \mu_e)^2 B^2}{(n_h \mu_h+n_e \mu_e)^2+(n_h-n_e)^2 (\mu_h \mu_e)^2 B^2},
\end{equation}
where $n_h(n_e)$ and $\mu_h(\mu_e)$ represent the hole (electron) concentration and hole (electron) mobility, respectively. Figures. 4(c) and 4(d) show the obtained concentration and mobility, respectively. At 3 K, $n_e=1.06\times10^{21}cm^{-3}$, $n_h=0.88\times10^{21}cm^{-3}$, $\mu_e=4.47\times10^3cm^2V^{-1}s^{-1}$ and $\mu_h=7.07\times10^3cm^2V^{-1}s^{-1}$. The ratio $n_h/n_e\approx0.83$ indicates the compensation of carriers in HoSb. From the field dependent resistivity
\begin{equation}\label{equ3}
\rho (B)=\frac{(n_h \mu_h+n_e \mu_e)+(n_h \mu_e+n_e \mu_h)\mu_h \mu_e B^2}{e (n_h \mu_h+n_e \mu_e)^2+e (n_h-n_e)^2 (\mu_h \mu_e)^2 B^2},
\end{equation}
the relation MR=$\mu_e\mu_h$\emph{B}$^2$ can be obtained for the perfect electron-hole compensation ($n_e$=$n_h$). Consequently, high mobility will lead to the large and unsaturated MR. If the ratio $n_h/n_e$ deviates from 1 slightly, MR will deviate from quadratic behavior slightly and remain unsaturated up to a high field\cite{PhysRevB.97.085137}. The XMR in HoSb originates from the electron-hole compensation and high mobility.

\begin{figure*}[htbp]
\centering
\includegraphics[width=0.7\textwidth]{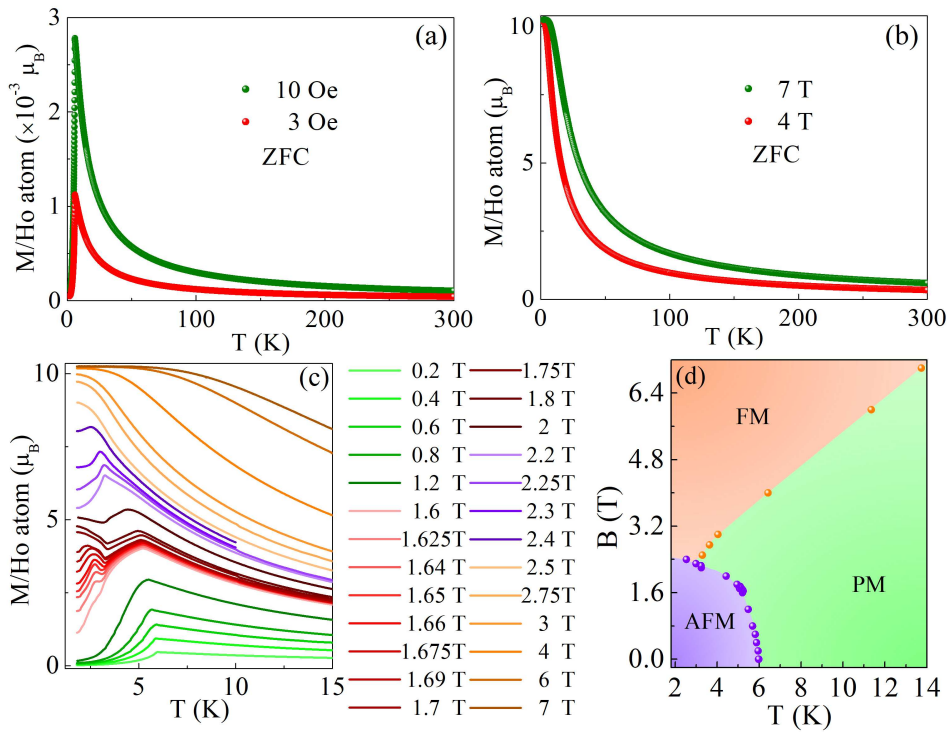}
\caption{Magnetic properties of HoSb (Sample 4) with the magnetic field parallel to the direction of [001]. (a) Temperature dependence of magnetization at low field 3 Oe and 10 Oe. (b) Temperature dependence of magnetization at high field 4 T and 7 T. (c) Temperature dependence of magnetization at various fields. (d) The magnetic phase diagram of HoSb.}
\end{figure*}

The field dependent MR has been examined to investigate the origin of the kink. As shown in Fig. 4(e), although the curve can be roughly fitted by MR$\propto$\emph{B}$^n$ and the obtained $n$=1.996 is close to 2, the fit is not perfect and the kink can be seen clearly as that in the Hall data. However, the data in high field can be well fitted (the inset of Fig. 4(e), $n$=1.84), indicating that the kink may originate from some transitions under low magnetic field. Then the field dependent magnetization was measured to examine the effect of the magnetic structure. As shown in Figs. 4(f) and 4(g), the magnetic moments of HoSb adopt MnO-type antiferromagnetic arrangement at low temperature, and the increase of field makes the arrangement change to HoP-type magnetic structure (manifested as the kink in the magnetization curve) and then to FM state\cite{PhysRev.131.922,brun1980quadropolar}. In the process of the transitions, the emergent disorder reduces the relaxation time $\tau$. Since $\mu=e\tau/m^*$, the decrease of relaxation time will cause the decrease of mobility, resulting in the formation of the kinks in the field dependent Hall resistivity and MR.

Figure 5 shows the study on the magnetic properties of HoSb. The magnetization curves exhibit a transition from PM to AFM at low temperature and low field (Fig. 5(a)). The transition temperature of 5.9 K is slightly larger than the value obtained from resistivity, which may be from the difference of samples. Under a high field, the AFM order is suppressed and the phase changes from PM to FM (Fig. 5(b)). The detailed measurements of the temperature-dependent magnetization at different fields have been performed to map the magnetic phase diagram. As shown in Fig. 5(c), the N$\acute{e}$el temperature decreases with the increase of field. In addition, an unexpected cusp at the temperature below $T_N$ has been found in a specific range of field, which was not reported in previous studies. Such an anomaly has also been observed in CuMnSb and attributed to the canted AFM structure without uniform magnetic moment\cite{PhysRevMaterials.2.054413}. Figure 5(d) plots the magnetic field and temperature phase diagram of HoSb, in which three regimes (PM, AFM, and FM) can be distinguished. With the increase of magnetic field, the AFM order is gradually suppressed and FM order gradually appears.

\section{Discussions}

Since the temperature $T_i$ in other nonmagnetic XMR materials is nearly field-independent\cite{tafti2015resistivity,PhysRevB.94.041103}, the observation of field-dependent $T_i$ in HoSb is interesting. Previously, the XMR and resistivity plateau in LaSb were attributed to the field-induced
metal-insulator transition and the saturation of insulating bulk resistance induced by the metallic surface conductance, respectively\cite{tafti2015resistivity}. Later, the possibility of field-induced metal-insulator transition was excluded\cite{tafti2016temperature,PhysRevB.96.125112}. Guo \emph{et al.} pointed out that the resistivity plateau comes from the nearly invariable carrier concentration and mobility at low temperature in compensated semimetals\cite{PhysRevB.93.235142}. Han \emph{et al.} suggested that the resistivity plateau originates from the temperature-insensitive resistivity at zero field if the MR can be scaled by Kohler's rule\cite{PhysRevB.96.125112}. However, for a two-component system that satisfies $n_e=n_h$ and $\mu_e=\mu_h$, the Kohler's rule can be easily derived from the two band model (Eq. 3). The temperature-insensitive resistivity at zero field is related to the nearly invariable carrier concentration and mobility at low temperature.

Different from nonmagnetic LaSb/LaBi/YSb and paramagnetic TmSb/PrSb, HoSb is an antiferromagnet and the magnetic structure can be changed by field. The increased disorder in the process of phase transitions reduces the mobility and leads to the kinks in $\rho(B)$ and $\rho_{xy}(B)$. Under different fields, the temperature where the FM phase appears is different. Meanwhile, the resistivity plateau always starts in the FM state. This indicates that the change of mobility induced by phase transitions has affected the emergence of resistivity plateau. It is suggested that the field dependence of $T_i$ is related to the field-dependent mobility in phase transitions.

\section{Summary}

In summary, single crystals of HoSb are grown and the magnetotransport properties have been studied. The unusual magnetotransport properties, including the \emph{field-dependent resistivity plateau} and the \emph{kinks in $\rho(B)$ and $\rho_{xy}(B)$ curves}, can be attributed to the magnetic phase transitions. HoSb also exhibits XMR and SdH oscillation at low temperature and high field. The FFT spectra of SdH oscillations reveal the split of Fermi surfaces. The nontrivial Berry phase of the electron bands indicates that HoSb may have topologically nontrivial electronic structure under the field. The electron-hole compensation and high mobility are suggested to be responsible for the observation of XMR.

\section{Acknowledgments}

We thank Hechang Lei and Lingxiao Zhao for helpful discussions. This work is supported by the National Natural Science Foundation of China (No.11574391), the Fundamental Research Funds for the Central Universities, and the Research Funds of Renmin University of China (No. 14XNLQ07, No. 18XNLG14).

\bibliography{bibtex}
\end{document}